\documentclass[aps,prd,nofootinbib,amsmath,amssymb,twocolumn,superscriptaddress,10pt]{revtex4}%superscriptaddress,showpacs,

\pdfoutput=1

\usepackage{graphicx}
\usepackage{dcolumn}
\usepackage{bm}
\usepackage{amssymb}
\usepackage{latexsym}
\usepackage{booktabs}
\usepackage{amsmath}
\usepackage{multirow}
\usepackage[colorlinks=true, linkcolor=red, citecolor=blue]{hyperref}

\newcommand{\be}{\begin{equation}}
\newcommand{\ee}{\end{equation}}
\newcommand{\bq}{\begin{eqnarray}}
\newcommand{\eq}{\end{eqnarray}}

\usepackage[usenames,dvipsnames]{xcolor}

\DeclareMathAlphabet\mathbfcal{OMS}{cmsy}{b}{n}
\definecolor{darkgreen}{cmyk}{0.85,0.2,1.00,0.2}
\definecolor{purple}{cmyk}{0.5,1.0,0,0}

\def\barray{\begin{array}}
\def\earray{\end{array}}
\def\be{\begin{equation}}
\def\ee{\end{equation}}
\def\ben{\begin{equation} \nonumber}
\def\een{\end{equation}}
\def\ban{\begin{eqnarray*}}
\def\ean{\end{eqnarray*}}
\def\ba{\begin{eqnarray}}
\def\ea{\end{eqnarray}}

\def\({\left(}
\def\){\right)}

\begin{document}

\title{Measuring growth index in a universe with massive neutrinos: A revisit of the general relativity test with the latest observations}

\author{Ming-Ming Zhao}
%\email{zhaomingmingsp@163.com}
\affiliation{Department of Physics, College of Sciences,
Northeastern University, Shenyang 110004, China}
\author{Jing-Fei Zhang}
%\email{jfzhang@mail.neu.edu.cn}
\affiliation{Department of Physics, College of Sciences,
Northeastern University, Shenyang 110004, China}
\author{Xin Zhang\footnote{Corresponding author}}
\email{zhangxin@mail.neu.edu.cn}
\affiliation{Department of Physics, College of Sciences,
Northeastern University, Shenyang 110004, China}
\affiliation{Center for High Energy Physics, Peking University, Beijing 100080, China}

\begin{abstract}

We make a consistency test for the general relativity (GR) through measuring the growth index $\gamma$ in a universe with massive (sterile/active) neutrinos. We employ the redshift space distortion measurements to do the analysis. To constrain other cosmological parameters, we also use other cosmological measurements, including the Planck 2015 cosmic microwave background temperature and polarization data, the baryon acoustic oscillation data, the type Ia supernova JLA data, the weak lensing galaxy shear data, and the Planck 2015 lensing data. In a universe with massive sterile neutrinos, we obtain $\gamma=0.624^{+0.055}_{-0.050}$, with the tension with the GR prediction $\gamma=0.55$ at the 1.48$\sigma$ level, showing that the consideration of sterile neutrinos still cannot make the true measurement of $\gamma$ be well consistent with the GR prediction. In a universe with massive active neutrinos, we obtain $\gamma=0.663\pm0.045$ for the normal hierarchy case, $\gamma=0.661^{+0.044}_{-0.050}$ for the degenerate hierarchy case, and $\gamma=0.668^{+0.045}_{-0.051}$ for the inverted hierarchy case, with the tensions with GR all at beyond the 2$\sigma$ level. We find that the consideration of massive active neutrinos (no matter what mass hierarchy is considered) almost does not influence the measurement of the growth index $\gamma$.

\end{abstract}

\maketitle

\section{Introduction}
The current astronomical observations have indicated that the universe is undergoing an accelerated expansion \cite{Riess:1998cb,Perlmutter:1998np,Tegmark:2003ud,Eisenstein:2005su,Spergel:2003cb}. To explain this accelerated expansion, in the context of general relativity (GR), the so-called dark energy (DE), an unknown component with negative pressure, is proposed \cite{Sahni:1999gb,Carroll:2000fy,Peebles:2002gy,Padmanabhan:2002ji,Copeland:2006wr}. On the other hand, the modification of gravity (MG) can also account for the accelerated expansion by mimicking the behavior of DE within GR for the whole expansion history at the background level \cite{Tsujikawa:2010zza,Clifton:2011jh,Ade:2015rim}. Both of them can in principle describe the same expansion, but they are different in nature. To distinguish between MG and GR, the precise large-scale structure (LSS) measurements are required because they have different histories of growth of structure.

A way to describe the growth of scalar (density) perturbations in non-relativistic matter component (cold dark matter and baryons) is provided by the parametrization $f(a)=\Omega_{\rm m}(a)^{\gamma}$, proposed in Ref.~\cite{Wang:1998gt}, where $f(a)\equiv d\ln\delta(a)/d\ln a$ is the growth rate for linear perturbations, $\Omega_{\rm m}(a)=\Omega_{\rm m}H_{0}^{2}H(a)^{-2}a^{-3}$ is the fractional matter density, and $\gamma$ is called the growth index. Both of the growth index and the evolution of matter density depend on the specific model (for details see the latest review \cite{Polarski:2016ieb}). For dark energy models with slowly varying equation of state, within GR, an approximation of $\gamma\approx0.55$ is derived. For example, based on the $\Lambda$CDM model, $\gamma=6/11\approx 0.545$ is given \cite{Linder:2005in}. However, for MG models, different theoretical values of $\gamma$ are derived; e.g., for the Dvali-Gabadadze-Porrati (DGP) model, $\gamma\approx0.68$ is obtained~\cite{Nojiri:2003ft,Capozziello:2003tk,Lue:2004rj}.

The growth index in a cosmological model can be constrained by using the redshift space distortion (RSD) observation. RSD is a significant probe for the growth of structure, which provides an important way of measuring the growth rate $f(z)$ at various redshifts. In practice, RSD measures the product of $f(z)$ and $\sigma_8(z)$, namely, $f(a)\sigma_8(a)=d\sigma_8(a)/d\ln a$, where $\sigma_8(z)$ is the root-mean-square mass fluctuation in a sphere of radius $8h^{-1}$ at the redshift $z$. However, using RSD to constrain the growth index (based on the $\Lambda$CDM model), it is found that there is a deviation of the $\gamma$ value from the GR's theoretical prediction of $\gamma\approx0.55$ at the 2--3$\sigma$ confidence level (see, e.g., Ref.~\cite{Xu:2013tsa}). Recently, Gil-Marin et al. \cite{Gil-Marin:2016wya} used the latest BOSS CMASS and LOWZ DR12 measurements combined with the Planck 2015 temperature and polarization spectra to constrain $\gamma$, and they obtained $\gamma=0.719^{+0.080}_{-0.072}$, which reveals a beyond $2.5\sigma$ tension with the GR prediction. The similar situation can be found in the previous studies \cite{Samushia:2012iq,Beutler:2013yhm}.

It can be noticed that the measurements of $\gamma$ have been always higher than the GR prediction. That is to say, the actual observed growth of structure is faster than that predicted by GR. One way to reconcile them is to consider massive (active or sterile) neutrinos in the cosmological model, since the free-streaming property of neutrinos could help suppress the growth of structure on small scales. In 2014, two of the authors of the present paper (Jing-Fei Zhang and Xin Zhang) and another collaborator (Yun-He Li) \cite{Zhang:2014lfa} considered this scheme, and they found that if massive sterile neutrinos are involved in the cosmological model, then the constraint value of $\gamma$ and the theoretical prediction of $\gamma=0.55$ will become well consistent.

In Ref.~\cite{Zhang:2014lfa}, Zhang, Li, and Zhang used the RSD data in combination with other observations (at that time) including the cosmic microwave background (CMB) anisotropy data from the Planck 2013 temperature spectrum \cite{Ade:2013zuv} and the WMAP 9-yr polarization data \cite{wmap9}, baryon acoustic oscillations (BAO) measurements from the 6dFGS \cite{6df}, SDSS DR7 \cite{sdss7}, WiggleZ \cite{wigglez}, and BOSS DR11 \cite{boss} surveys, the Hubble constant $H_{0}$ measurement with the value of $73.8\pm2.4\, \rm km/s/Mpc$ \cite{h0}, the Planck Sunyaev-Zeldovich cluster counts data \cite{tsz}, and the cosmic shear data from CFHTLenS survey \cite{shear}, to constrain $\gamma$, and they obtained $\gamma=0.584^{+0.047}_{-0.048}$, well consistent with the prediction value of GR of $\gamma\approx0.55$. See Refs.~\cite{Ade:2015xua,Viel:2005qj,Lesgourgues:2006nd,Viel:2006kd,Lesgourgues:2014zoa,Palazzo:2013me,Leistedt:2014sia,Ko:2014bka,Archidiacono:2014nda,Li:2014dja,An:2014bik,Wyman:2013lza,
Hamann:2013iba,DiValentino:2013qma,Zhang:2014nta,Li:2015poa,Feng:2017mfs} for other previous works discussing the issue of using massive sterile neutrinos to relieve tensions among cosmological observations.

However, it should be pointed out that it is the time to revisit this issue with the latest cosmological observations. In the past years, numerous more accurate data were released, which would update the previous results derived in Ref.~\cite{Zhang:2014lfa} and even would change the conclusive statements.

In this paper, we will revisit the study of the constraints on the growth index, based on the $\Lambda$CDM cosmology with massive (sterile/active) neutrinos, using the latest cosmological measurements, including the Planck 2015 temperature and polarization power spectra and the latest RSD data. Moreover, since the lensing observations including the weak lensing and the CMB lensing can capture the effects of massive neutrinos on the matter power spectra, they can provide useful constraint on the neutrino mass. In addition, the growth index is related to not only the structure's growth, but also the expansion of the universe, and thus the independent geometric observations such as BAO and type of Ia supernova (SN) are also needed. We will use these latest observations to study the measurement of the growth index $\gamma$.

The paper is arranged as the following. In Sec.~\ref{data}, we introduce the method and observational data used in this paper. In Sec.~\ref{sec:results}, we report the results of the consistency test of GR. In Sec.~\ref{sec:discussion}, we will make a conclusion for this work.

\section{Method and data}\label{data}

In this paper, we place constraints on the growth index $\gamma$ in the $\Lambda$CDM cosmology with massive (sterile/active) neutrinos with the latest observations. Within GR, as long as the equation-of-state parameter of DE is slowly varying, the theoretical predictions of $\gamma$ for DE models are almost the same, i.e., $\gamma\approx 0.55$. Thus, in this paper, we only consider the $\Lambda$CDM cosmology.

For the base $\Lambda$CDM model, there are six base parameters, which are the baryon density $\Omega_{\rm b}h^{2}$, the cold dark matter density $\Omega_{\rm c}h^{2}$, the ratio of the angular diameter distance to the sound horizon at last scattering $\theta_\ast$, the reionization optical depth $\tau$, and the amplitude $A_{s}$ and the tilt $n_{s}$ of the primordial scalar fluctuations.

To constrain the growth index $\gamma$, we use the parametrization $f(a)=\Omega_{\rm m}(a)^{\gamma}$ to describe the density perturbations in the $\Lambda$CDM cosmology, and thus we introduce an extra parameter $\gamma$ into the model. We use the RSD measurements of $f(z_{\rm eff})\sigma_8(z_{\rm eff})$ to set constraints on $\gamma$. We follow the procedure of Sec.~9.1 in Ref.~\cite{Beutler:2013yhm} to include $\gamma$ as an additional parameter. Here, it is helpful to briefly describe how the parameter product $f_\gamma(z_{\rm eff})\sigma_{8,\gamma}(z_{\rm eff})$ is derived in the theoretical calculations by the following two steps: (i) Since in this description the value of $\sigma_8(z_{\rm eff})$ depends on $\gamma$, we have to recalculate this value by using the parametrization of $f_\gamma(a_{\rm eff})=\Omega_{\rm m}(a_{\rm eff})^{\gamma}$. We first calculate the growth factor, $D(a_{\rm eff})=\exp\left[-\int_{a_{\rm eff}}^1 da'f(a')/a'\right]$,
%\begin{equation}
%D(a_{\rm eff})=\exp\left[-\int_{a_{\rm eff}}^1 da'f(a')/a'\right],
%\end{equation}
where $a_{\rm eff}$ is the scale factor at the effective redshift $z_{\rm eff}$. Then, we derive $\sigma_{8,\gamma}(z_{\rm eff})$ by the extrapolation from the matter dominated epoch to the effective redshift, $\sigma_{8,\gamma}(z_{\rm eff})={D_\gamma(z_{\rm eff})\over D(z_{\rm hi})}\sigma_8(z_{\rm hi})$,
%\begin{equation}
%\sigma_{8,\gamma}(z_{\rm eff})={D_\gamma(z_{\rm eff})\over D(z_{\rm hi})}\sigma_8(z_{\rm hi}),
%\end{equation}
where $\sigma_8(z_{\rm hi})$ is calculated at $z_{\rm hi}=50$ which is in the deep matter-dominated regime, where $f(z)\approx 1$. (ii) We calculate the growth rate by using the parametrization $f_\gamma(z_{\rm eff})=\Omega_{\rm m}(z_{\rm eff})^{\gamma}$. Thus, now, we can obtain the parameter product $f_\gamma(z_{\rm eff})\sigma_{8,\gamma}(z_{\rm eff})$ in the numerical calculations.

If we further consider massive neutrinos in cosmology, we need to add the total neutrino mass $\sum m_{\nu}$ for the case of active neutrino and the effective mass $m_{\nu,\rm sterile}^{\rm eff}$ and the effective number of relativistic species $N_{\rm eff}$ for the case of sterile neutrino.

We make our analysis by employing several important cosmological probes. We use the Planck 2015 full temperature and polarization power spectra at $2\leq~\ell~\leq~2900$. We refer to this dataset as ``Planck TT,TE,EE'' (note that we do not use ``+lowP'', as the Planck collaboration used, for simplicity). In addition to the CMB dataset described above, we consider the combination with the following cosmological measurements:

\begin{itemize}
 \item The BAO data: We use the BAO measurements from the 6dFGS ($z=0.1$) \cite{6df}, SDSS-MGS ($z=0.15$) \cite{Ross:2014qpa}, LOWZ ($z=0.32$) and CMASS ($z=0.57$) DR12 samples of BOSS \cite{Alam:2016hwk}. Note that we exclude the BOSS DR12 results from BAO likelihood when the RSD measurements of DR12 are used in the data combination in this paper.
     \item The SN data: For the type Ia supernova observation, we adopt the ``JLA" sample, compiled from the SNLS, SDSS, and the samples of several low-redshift SN data \cite{Betoule:2014frx}.
              \item The RSD data: We employ RSD measurements at $11$ redshifts, which are 6dFGS ($z=0.067$)~\cite{RSD6dF}, 2dFGS ($z=0.17$)~\cite{RSD2dF}, WiggleZ ($z=0.22, 0.41, 0.60$, and $0.78$)~\cite{RSDwigglez}, SDSS LRG DR7 ($z=0.25$ and $z=0.37$)~\cite{RSDsdss7}, BOSS CMASS DR12 ($z=0.57$) and LOWZ DR12 ($z=0.32$)~\cite{Gil-Marin:2016wya}, and VIPERS ($z=0.80$)~\cite{RSDvipers} samples.
                   \item The WL and CMB lensing data: We use the cosmic shear measurement of weak lensing from the CFHTLenS survey, and we apply the ``conservative'' cuts for the shear data according to the recipe of Ref.~\cite{Heymans:2013fya}. We denote the dataset of shear measurement (weak lensing) as ``WL'' in this paper. We also use the CMB lensing power spectrum from the Planck 2015 lensing measurement \cite{Ade:2015zua}, which is denoted as ``lensing'' in this paper.

                   \end{itemize}

The analysis is done with the latest version of the publicly available Markov-Chain Monte Carlo package \texttt{CosmoMC} \cite{Lewis:2002ah}, with a convergence diagnostic based on the Gelman and Rubin statistics.

In this paper, tensions between different observations for some cosmological parameters will occasionally be mentioned, so it is helpful to clearly describe how to estimate the degree of tension between two observations for some parameter in this place. Assume that, for a parameter $\xi$, we have its 68\% confidence level ranges $\xi\in [\xi_1-\sigma_{\rm 1, low}, ~\xi_1+\sigma_{\rm 1, up}]$ from an observation (O1) and $\xi\in [\xi_2-\sigma_{\rm 2, low}, ~\xi_2+\sigma_{\rm 2, up}]$ from another observation (O2). The statement that ``the tension between O1 and O2 is at the $a\sigma$ level'' means that we have $a=(\xi_2-\xi_1)/\sqrt{\sigma_{\rm 2, low}^2+\sigma_{\rm 1, up}^2}$ for the case $\xi_2>\xi_1$, and vice versa. In this work, we estimate the degree of tension between different observations by this simple way.

\section{Results}\label{sec:results}

\subsection{Sterile neutrinos and growth index}

\begin{figure}
\begin{center}
\includegraphics[scale=0.8, angle=0]{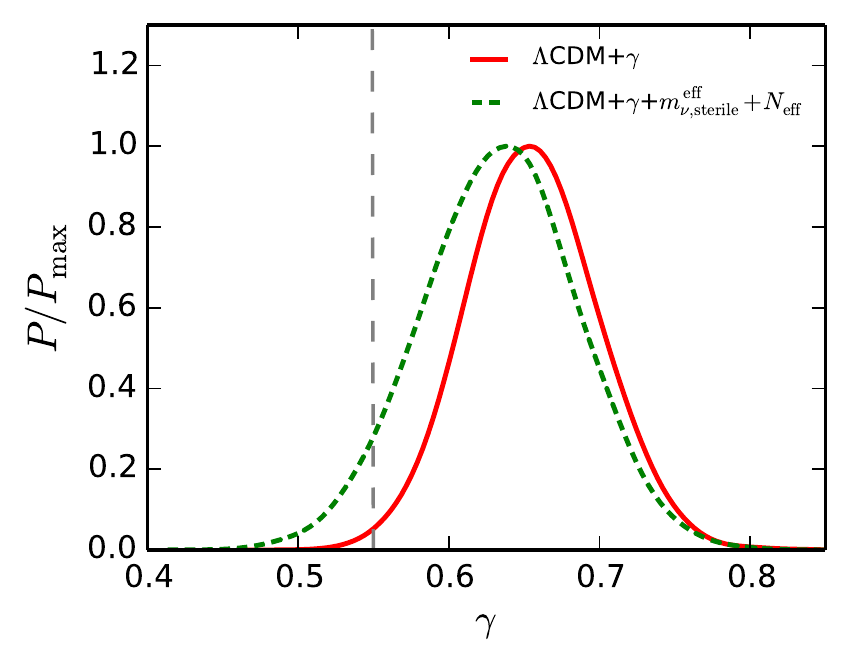}
\includegraphics[scale=0.8, angle=0]{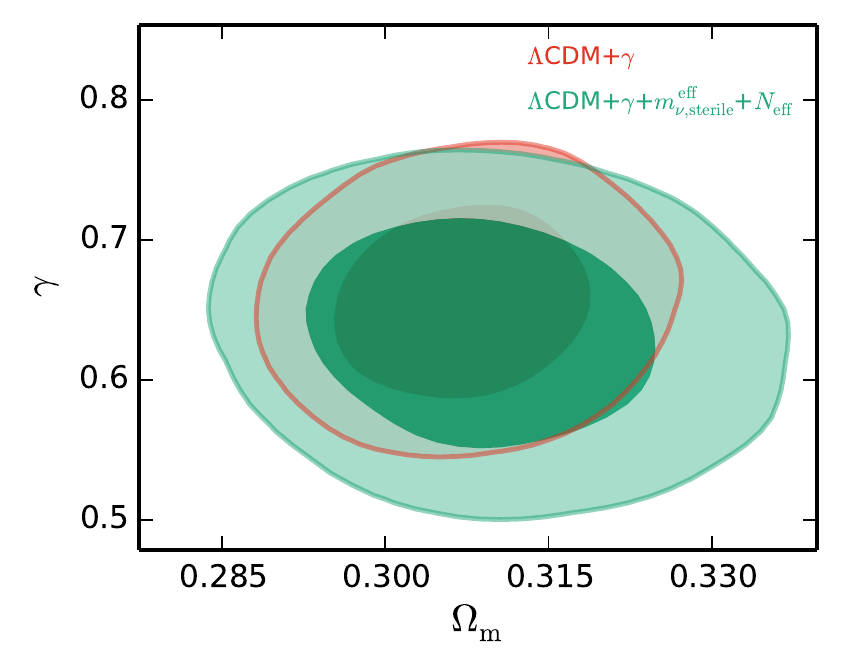}
\caption{One-dimensional and two-dimensional joint, marginalized constraints on the $\Lambda$CDM+$\gamma$ model and the $\Lambda$CDM+$\gamma$+$m_{\nu,\rm sterile}^{\rm eff}$+$N_{\rm eff}$ model from the data combination of Planck TT,TE,EE+BAO+SN+WL+RSD+lensing. The top panel shows the one-dimensional distribution of $\gamma$, and the bottom panel shows the two-dimensional distribution contours (68\% and 95\% confidence level) in the $\Omega_{m}-\gamma$ plane.}
\label{fig:gamma1}
\end{center}
\end{figure}

We constrain the growth index in the $\Lambda$CDM cosmology with sterile neutrinos by using the data combination of Planck TT,TE,EE+BAO+SN+RSD+WL+lensing. We show the constraint results in  Fig.~\ref{fig:gamma1} and Table~\ref{tab1}. Here, with the purpose of visually showing the effect of sterile neutrinos on the constraints of the growth index, we also perform an analysis for the $\Lambda$CDM+$\gamma$ model (without sterile neutrinos), to make a comparison, and the detailed results are presented in Fig.~\ref{fig:gamma1} and Table~\ref{tab1}.

\begin{table}
\caption{\label{tab1} Fitting results for the $\Lambda$CDM+$\gamma$ and $\Lambda$CDM+$\gamma$+$N_{\rm eff}$+$m_{\nu,\rm sterile}^{\rm eff}$ models from the data combination Planck TT,TE,EE+BAO+SN+WL+RSD+lensing. Here, we quote the $\pm 1\sigma$ errors, but for the parameters $m_{\nu,\rm sterile}^{\rm eff}$ and $N_{\rm eff}$, we quote the 95\% CL upper limits.}

\begin{tabular}{ccccccccc}
\hline % \multicolumn{1}{c}{data}&&\multicolumn{3}{c}{Planck TT,TE,EE+BAO+SN+RSD+WL+lensing+$H_{0}$}\\
          %\cline{1-1}\cline{3-5}\cline{7-9}

      % Model&$\Lambda$CDM+$\gamma$ &$\Lambda$CDM+$\gamma$+$\sum m_{\nu}$ (normal)&$\Lambda$CDM+$\gamma$+$\sum m_{\nu}$(degenerate)&$\Lambda$CDM+$\gamma$+$\sum m_{\nu}$ (inverted)\\
       Model&$\Lambda$CDM+$\gamma$&$\Lambda$CDM+$\gamma$+$m_{\nu,\rm sterile}^{\rm eff}$+$N_{\rm eff}$\\
\hline

$\Omega_{\rm b}h^2$&$0.02231\pm0.00015$&$0.02247\pm0.00016$\\
$\Omega_{\rm c}h^2$&$0.1184\pm0.0012$&$0.1167^{+0.0035}_{-0.0026}$\\
$100\theta_{\rm MC}$&$1.04089\pm0.00031$&$1.0409^{+0.00035}_{-0.00031}$\\
$\tau$&$0.058^{+0.013}_{-0.014}$&$0.071^{+0.014}_{-0.015}$\\
$n_s$&$0.9668\pm0.0044$&$0.971^{+0.0047}_{-0.0066}$\\
${\rm{ln}}(10^{10}A_s)$&$3.047^{+0.024}_{-0.025}$&$3.074^{+0.026}_{-0.030}$\\
$\gamma$&$0.656^{+0.042}_{-0.046}$&$0.624^{+0.055}_{-0.050}$\\
$\Omega_m$&$0.3073\pm0.0073$&$0.3094^{+0.0113}_{-0.0093}$\\
$m_{\nu,\rm sterile}^{\rm eff}$ [eV]&$-$&$<0.743$\\
$N_{\rm eff}$&$-$&$<3.33$\\

\hline
\end{tabular}
\end{table}

Figure~\ref{fig:gamma1} displays the one-dimensional posterior distribution for $\gamma$ in the top panel and the two-dimensional, joint, marginalized constraints (68\% and 95\% confidence level) in the $\gamma-\Omega_{\rm m}$ plane in the bottom panel. First, let us have a look at the constraint results of $\gamma$ (see Table~\ref{tab1}). We obtain $\gamma=0.656^{+0.042}_{-0.046}$ for the $\Lambda$CDM+$\gamma$ model and $\gamma=0.624^{+0.055}_{-0.050}$ for the $\Lambda$CDM+$\gamma$+$m_{\nu,\rm sterile}^{\rm eff}$+$N_{\rm eff}$ model, which indicates that the tensions of $\gamma$ with the GR prediction are at the $2.30\sigma$ level and the $1.48\sigma$ level for the two cases, respectively. We find that, in the $\Lambda$CDM+$\gamma$+$N_{\rm eff}$+$m_{\nu,\rm sterile}^{\rm eff}$ model, the tension is milder than that of the model without sterile neutrinos. In the top panel of Fig.~\ref{fig:gamma1}, the one-dimensional posterior distributions of $\gamma$ show that the consideration of sterile neutrinos indeed leads the fit value of $\gamma$ to be more consistent with the GR theoretical value of $\gamma=0.55$, presented by the grey dotted line. According to the fitting results, we can clearly see that once a light sterile neutrino species is considered in the universe, a smaller value of $\gamma$ is indeed derived, but it is not enough to lead the true measurement of $\gamma$ to be well consistent with the GR prediction.

The tension with GR might be related to the matter density $\Omega_{\rm m}$ in the $\Lambda$CDM model fitting to the CMB data \cite{Gil-Marin:2016wya}. The change in $\gamma$ also depends on a change in determination of $\Omega_{\rm m}$ using the CMB and other data. We show the contours in the $\gamma-\Omega_{\rm m}$ plane in the bottom panel of Fig.~\ref{fig:gamma1}. We find that the correlation between $\gamma$ and $\Omega_{\rm m}$ is weak for the both models. We also see that once sterile neutrinos are considered in the model, the value of $\Omega_{\rm m}$ is increased and the value of $\gamma$ is somewhat lowered, although the whole range of the contour is amplified due to the addition of extra two parameters.

We also compare our results with those of the previous study~\cite{Zhang:2014lfa}. In Ref.~\cite{Zhang:2014lfa}, it is found that when a sterile neutrino species is considered in the model, the tension with GR will be at the less than 1$\sigma$ level. But, in this study, we find that even though sterile neutrinos are considered, the tension with GR will still be at the more than 1$\sigma$ level (but less than 2$\sigma$ level). In Ref.~\cite{Zhang:2014lfa}, for the constraint on the parameters of sterile neutrino, the authors obtain $N_{\rm eff}=3.62^{+0.26}_{-0.42}$ and $m_{\nu,\rm sterile}^{\rm eff}=0.48^{+0.11}_{-0.14}$ eV, indicating the preference for $\Delta N_{\rm eff}\equiv N_{\rm eff}-3.046>0$ at the 1.4$\sigma$ level and for nonzero mass of sterile neutrino at the 3.4$\sigma$ level. But, in the present study, we only obtain upper limits on both $N_{\rm eff}$ and $m_{\nu,\rm sterile}^{\rm eff}$, namely, $N_{\rm eff}<3.33$ and $m_{\nu,\rm sterile}^{\rm eff}<0.743$ eV.
%, which means that we find no hint of the existence of light sterile neutrinos.
Recently, the neutrino oscillation experiments by Daya Bay and MINOS collaborations \cite{Adamson:2016jku}, as well as the cosmic ray experiment by the IceCube collaboration \cite{TheIceCube:2016oqi}, found no evidence for a massive sterile neutrino species, in good consistency with our present result.

We also wish to simply address the issue of other tensions. Including sterile neutrinos in a universe can enhance the fit value of $H_{0}$ (due to the positive correlation between $N_{\rm eff}$ and $H_0$ existing in the CMB fit), and hence reconcile the tension between Planck and the direct measurement of the Hubble constant ($H_{0}=73.00\pm1.75 ~\rm km  ~s^{-1} ~Mpc^{-1}$ \cite{Riess:2016jrr}) \cite{Zhao:2017urm,Feng:2017nss,Dvorkin:2014lea,Zhang:2014dxk,Battye:2013xqa,Wyman:2013lza,Hamann:2013iba}. We present the constraint values of $H_{0}$ and tensions with the direct measurement of the Hubble constant for the $\Lambda$CDM+$\gamma$ model and the $\Lambda$CDM+$\gamma$+$N_{\rm eff}$+$m_{\nu,\rm sterile}^{\rm eff}$ model in Table~\ref{tab2}. From the table, we can see that $H_{0}=67.84\pm0.55\,\rm km\,s^{-1}\,Mpc^{-1}$ for the $\Lambda$CDM+$\gamma$ model and $H_{0}=68.42^{+0.67}_{-1.10}\,\rm km\,s^{-1}\,Mpc^{-1}$ for the $\Lambda$CDM+$\gamma$+$N_{\rm eff}$+$m_{\nu,\rm sterile}^{\rm eff}$ model, indicating that the tensions with the local determination of the Hubble constant are at the $2.8\sigma$ level and the $2.4\sigma$ level, respectively. This means that the consideration of sterile neutrinos offers only a marginal improvement for the issue of $H_0$ tension.

\begin{table}
\caption{\label{tab2} Fitting results of $H_{0}$ and their tensions with the direct measurement of the Hubble constant for the $\Lambda$CDM+$\gamma$ and $\Lambda$CDM+$\gamma$+$N_{\rm eff}$+$m_{\nu,\rm sterile}^{\rm eff}$ models from the data combination Planck TT,TE,EE+BAO+SN+WL+RSD+lensing.}

\begin{tabular}{|c|c|c|}
\hline % \multicolumn{1}{c}{data}&&\multicolumn{3}{c}{Planck TT,TE,EE+BAO+SN+RSD+WL+lensing+$H_{0}$}\\
          %\cline{1-1}\cline{3-5}\cline{7-9}

      % Model&$\Lambda$CDM+$\gamma$ &$\Lambda$CDM+$\gamma$+$\sum m_{\nu}$ (normal)&$\Lambda$CDM+$\gamma$+$\sum m_{\nu}$(degenerate)&$\Lambda$CDM+$\gamma$+$\sum m_{\nu}$ (inverted)\\
       Model&$\Lambda$CDM+$\gamma$&$\Lambda$CDM+$\gamma$+$m_{\nu,\rm sterile}^{\rm eff}$+$N_{\rm eff}$\\
\hline
$H_0$&$67.84\pm0.55$&$68.42^{+0.67}_{-1.10}$\\
\hline
tension&$2.8\sigma$&$2.4\sigma$\\
\hline
\end{tabular}
\end{table}

Next, we examine the tension between Planck and the recent observations of large-scale structure, mainly in terms of constraining the parameter $S_{8}\equiv\sigma_{8}\sqrt{\Omega_{\rm m}/0.3}$. In the last year, the tomographic weak gravitational lensing analysis of $\sim$ 450 $\rm deg^{2}$ from the Kilo Degree Surveys (KiDS-450) gave their latest cosmological parameter constraints, in which they obtained $S_{8}=0.745\pm0.039$ assuming a flat $\Lambda$CDM model \cite{Hildebrandt:2016iqg}. More recently, the cosmological result from a combined analysis of galaxy clustering and weak gravitational lensing, using 1321 ${\rm deg}^{2}$ of $griz$ imaging data from the first year of the Dark Energy Survey (DES Y1) was presented in Ref.~\cite{Abbott:2017wau}. They obtained $S_{8}=0.783^{+0.021}_{-0.025}$ for the $\Lambda$CDM cosmology. In the present work, we also calculated the $S_{8}$ values for the two models. In Table.~\ref{tab3}, we show our fit results. We obtain $S_{8}=0.813^{+0.010}_{-0.009}$ for the $\Lambda$CDM+$\gamma$ model and $S_{8}=0.795^{+0.022}_{-0.015}$ for the $\Lambda$CDM+$\gamma$+$N_{\rm eff}$+$m_{\nu,\rm sterile}^{\rm eff}$ model. According to our fit results, in the $\Lambda$CDM+$\gamma$ model, the tension with KiDS-450 is at the $1.7\sigma$ level and the tension with DES Y1 is at the $1.3\sigma$ level; in the $\Lambda$CDM+$\gamma$+$N_{\rm eff}$+$m_{\nu,\rm sterile}^{\rm eff}$ model, the tension is somewhat relieved, i.e., the tension with KiDS-450 is relieved to be at the $1.2\sigma$ level and the tension with DES Y1 is relieved to be at the $0.5\sigma$ level (see also Table.~\ref{tab3}).

\begin{table}
\caption{\label{tab3} Fitting results of $S_{8}$ and their tensions with KiDS-450 and DES Y1, respectively, for the $\Lambda$CDM+$\gamma$ and $\Lambda$CDM+$\gamma$+$N_{\rm eff}$+$m_{\nu,\rm sterile}^{\rm eff}$ models from the data combination Planck TT,TE,EE+BAO+SN+WL+RSD+lensing.}

\begin{tabular}{|c|c|c|}
\hline % \multicolumn{1}{c}{data}&&\multicolumn{3}{c}{Planck TT,TE,EE+BAO+SN+RSD+WL+lensing+$H_{0}$}\\
          %\cline{1-1}\cline{3-5}\cline{7-9}

      % Model&$\Lambda$CDM+$\gamma$ &$\Lambda$CDM+$\gamma$+$\sum m_{\nu}$ (normal)&$\Lambda$CDM+$\gamma$+$\sum m_{\nu}$(degenerate)&$\Lambda$CDM+$\gamma$+$\sum m_{\nu}$ (inverted)\\
       Model&$\Lambda$CDM+$\gamma$&$\Lambda$CDM+$\gamma$+$m_{\nu,\rm sterile}^{\rm eff}$+$N_{\rm eff}$\\
\hline
$S_{8}$&$0.813^{+0.010}_{-0.009}$&$0.795^{+0.022}_{-0.015}$\\
\hline
tension (KiDS-450)&$1.7\sigma$&$1.2\sigma$\\
\hline
tension (DES Y1)&$1.3\sigma$&$0.5\sigma$\\
\hline
\end{tabular}
\end{table}

\subsection{Active neutrinos and growth index}

\begin{figure}
\begin{center}
\includegraphics[scale=0.8, angle=0]{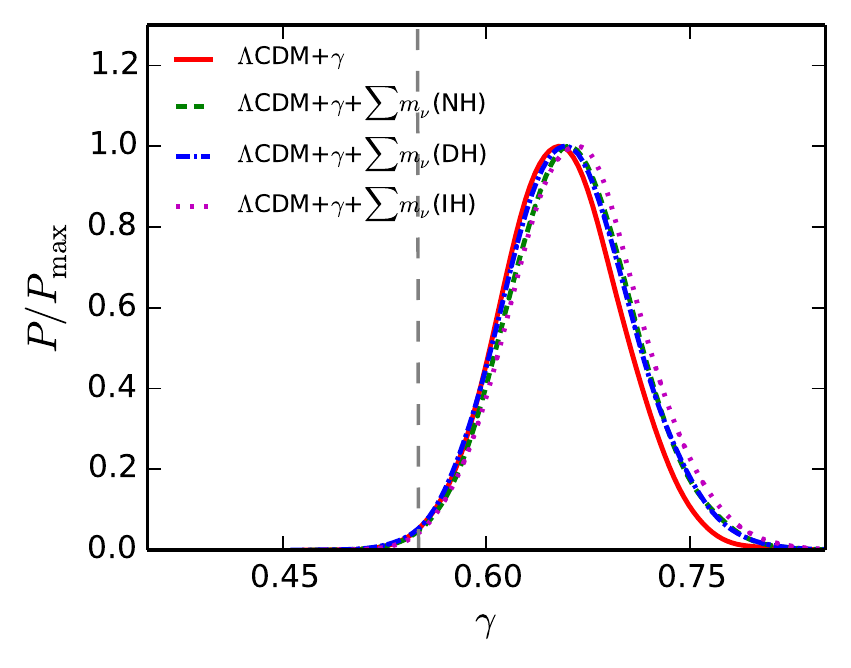}
\includegraphics[scale=0.8, angle=0]{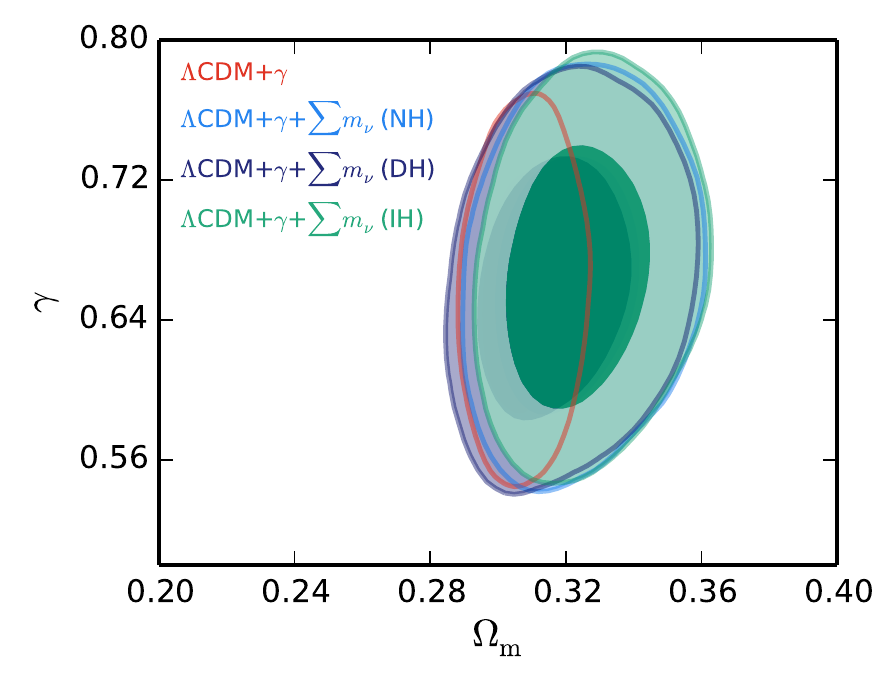}
\caption{One-dimensional and two-dimensional joint, marginalized constraints on the $\Lambda$CDM+$\gamma$ model and the $\Lambda$CDM+$\gamma$+$\sum m_{\nu}$ (normal, degenerate, and inverted hierarchies) models from the data combination of Planck TT,TE,EE+BAO+SN+WL+RSD+lensing. The top panel shows the one-dimensional distribution of $\gamma$, and the bottom panel shows the two-dimensional distribution contours (68\% and 95\% confidence level) in the $\Omega_{\rm m}-\gamma$ plane.}
\label{fig:gamma2}
\end{center}
\end{figure}

\begin{table*}
\caption{\label{tab4} Fitting results for the $\Lambda$CDM+$\gamma$ and $\Lambda$CDM+$\gamma$+$\sum m_{\nu}$ models (normal, degenerate, and inverted hierarchies) from the Planck TT,TE,EE+BAO+SN+WL+RSD+lensing data combination. Here, we quote the $\pm 1\sigma$ errors, but for the neutrino mass $\sum m_{\nu}$, we quote the 95\% CL upper limits.}
\centering
\begin{tabular}{ccccccccc}
%\hline  \multicolumn{1}{c}{data}&&\multicolumn{3}{c}{Planck TT,TE,EE+BAO+SN+RSD+WL+lensing}\\
          %\cline{1-1}\cline{3-5}\cline{7-9}
%\hline
\hline
       Model&$\Lambda$CDM+$\gamma$ &$\Lambda$CDM+$\gamma$+$\sum m_{\nu}$ (NH)&$\Lambda$CDM+$\gamma$+$\sum m_{\nu}$(DH)&$\Lambda$CDM+$\gamma$+$\sum m_{\nu}$ (IH)\\
\hline

$\Omega_{\rm b}h^2$&$0.02231\pm0.00015$&$0.02230\pm0.00015$&$0.02230\pm0.00015$&$0.02230\pm0.00015$\\
$\Omega_{\rm c}h^2$&$0.1184\pm0.0012$&$0.1184\pm0.0012$&$0.1184\pm0.0012$&$0.1184\pm0.0012$\\
$100\theta_{\rm MC}$&$1.04089\pm0.00031$&$1.04081\pm0.00031$&$1.04083\pm0.00031$&$1.04080\pm0.00031$\\
$\tau$&$0.058^{+0.013}_{-0.014}$&$0.069^{+0.015}_{-0.017}$&$0.066^{+0.016}_{-0.018}$&$0.071^{+0.015}_{-0.017}$\\
$n_s$&$0.9668\pm0.0044$&$0.9668\pm0.0044$&$0.9669^{+0.0043}_{-0.0044}$&$0.9669^{+0.0044}_{-0.0045}$\\
${\rm{ln}}(10^{10}A_s)$&$3.047^{+0.024}_{-0.025}$&$3.068^{+0.029}_{-0.031}$&$3.063^{+0.029}_{-0.034}$&$3.074^{+0.029}_{-0.031}$\\
$\Sigma m_\nu$[eV]&-&$<0.390$&$<0.374$&$<0.406$\\
$\gamma$&$0.656^{+0.042}_{-0.046}$&$0.663\pm0.045$&$0.661^{+0.044}_{-0.050}$&$0.668^{+0.045}_{-0.051}$\\
$\Omega_m$&$0.307\pm0.007$&$0.322^{+0.011}_{-0.016}$&$0.318^{+0.011}_{-0.018}$&$0.325^{+0.010}_{-0.016}$\\
$H_0$&$67.85\pm0.55$&$66.67^{+1.27}_{-0.85}$&$66.98^{+1.40}_{-0.94}$&$66.42^{+1.22}_{-0.80}$\\
$\sigma_8$&$0.808\pm0.009$&$0.787^{+0.023}_{-0.013}$&$0.792^{+0.025}_{-0.015}$&$0.782^{+0.021}_{-0.013}$\\

\hline
$\chi^{2}_{\rm min}$ & 13765.68&13764.51 &13764.65 & 13765.55 \\
\hline
\end{tabular}
\end{table*}

In this subsection, we constrain the growth index in a universe with massive active neutrinos. In this case, $N_{\rm eff}$ is fixed at 3.046 and $\sum m_\nu$ is freely varied within a prior range.

In this investigation, we also consider the mass splitting of neutrinos. The solar and reactor experiments measured $\Delta m_{21}^2\simeq 7.5\times 10^{-5}$ eV$^2$, and the atmospheric and accelerator beam experiments gave $|\Delta m_{31}^2|\simeq 2.5\times 10^{-3}$ eV$^2$, indicating that there are two possible mass orders, i.e., the normal hierarchy (NH) with $m_1<m_2\ll m_3$ and the inverted hierarchy (IH) with $m_3\ll m_1<m_2$. According to these two measured values, for the NH case, the neutrino mass spectrum can be written as $(m_1,m_2,m_3) = (m_1,\sqrt{m_1^2+\Delta m_{21}^2},\sqrt{m_1^2+|\Delta m_{31}^2|})$, in terms of a free parameter $m_1$; and for the IH case, the neutrino mass spectrum is expressed as $(m_1,m_2,m_3) = (\sqrt{m_3^2+|\Delta m_{31}^2|},\sqrt{m_3^2+|\Delta m_{31}^2|+\Delta m_{21}^2},m_3)$, in terms of $m_3$. We also consider the degenerate hierarchy (DH) case, i.e., $m_1=m_2=m_3=m$, where $m$ is a free parameter. Note that the input lower bound of $\sum m_\nu$ is $0.06$ eV, 0.10 eV, and $0$ for NH, IH, and DH, respectively. See Refs.~\cite{Huang:2015wrx,Wang:2016tsz,Vagnozzi:2017ovm,Capozzi:2017ipn,Xu:2016ddc,Gariazzo:2018pei} for details about the consideration of the mass hierarchies of neutrinos.

We use the same data combination, i.e., Planck TT,TE,EE+BAO+SN+WL+RSD+lensing, to do the analysis. The detailed fit results are shown in Table~\ref{tab4} and Fig.~\ref{fig:gamma2}. In Table~\ref{tab4}, for a direct comparison, we duplicate the results of the $\Lambda$CDM+$\gamma$ model here (see also Table~\ref{tab1}). We obtain $\gamma=0.663\pm0.045$ for the $\Lambda$CDM+$\gamma$+$\sum m_{\nu}$ (NH) model, $\gamma=0.661^{+0.044}_{-0.050}$ for the $\Lambda$CDM+$\gamma$+$\sum m_{\nu}$ (DH) model, and $\gamma=0.668^{+0.045}_{-0.051}$ for the $\Lambda$CDM+$\gamma$+$\sum m_{\nu}$ (IH) model. In the $\Lambda$CDM+$\gamma$ model, we have $\gamma=0.656^{+0.042}_{-0.046}$. So, we find that the consideration of neutrino mass (no matter what mass hierarchy is considered) does not influence the measurement of the growth index $\gamma$.

Figure~\ref{fig:gamma2} shows the one-dimensional posterior distributions for $\gamma$ (the top panel) and the two-dimensional marginalized contours in the $\Omega_{m}-\gamma$ plane (the bottom panel) for the $\Lambda$CDM+$\gamma$ and $\Lambda$CDM+$\gamma$+$\sum m_{\nu}$ models. From the top panel of Fig.~\ref{fig:gamma2}, we can clearly see that the posterior distribution curves are almost in coincidence. The tensions with the standard value of GR prediction $\gamma=0.55$ are still at the $2.51\sigma$, $2.22\sigma$, and $2.31\sigma$ level, for the NH, DH, and IH cases, respectively. In the bottom panel of Fig.~\ref{fig:gamma2}, we can see that, after the consideration of neutrino mass, the range of $\Omega_{\rm m}$ is enlarged (and the value of $\Omega_{\rm m}$ is also enhanced), and the range of $\gamma$ is nearly unchanged.

Next, we report the constraint results of the active neutrino mass in a universe with the density perturbation parametrized by the growth index $\gamma$. Using the Planck TT,TE,EE+BAO+SN+RSD+WL+lensing data combination, we obtain $\sum m_{\nu}<0.390$ eV for the $\Lambda$CDM+$\gamma$+$\sum m_{\nu}$ (NH) model, $\sum m_{\nu}<0.374$ eV for the $\Lambda$CDM+$\gamma$+$\sum m_{\nu}$ (DH) model, and $\sum m_{\nu}<0.406$ eV for the $\Lambda$CDM+$\gamma$+$\sum m_{\nu}$ (IH) model. These constraints are not the most stringent, compared to other recent upper limits of the neutrino mass such as reported in Refs.~\cite{Ade:2015xua,Zhang:2015uhk,Huang:2015wrx,Wang:2016tsz,Zhao:2016ecj,Guo:2017hea,Zhang:2017rbg,Couchot:2017pvz,Archidiacono:2017tlz,Caldwell:2017mqu,Doux:2017tsv,
Wang:2017htc,Chen:2017ayg,DiValentino:2017oaw,Lattanzi:2017ubx,Feng:2017usu}. This is because in this work we consider the neutrino mass in a cosmological model with extra parameter $\gamma$ (and also we use different observational data). In this work, we obtain the almost same values of $\chi_{\rm min}^2$ for the three mass hierarchy cases, indicating that the current cosmological observations still cannot diagnose the mass hierarchy of neutrinos in a universe with the density perturbation parametrized by the growth index $\gamma$.

\section{Conclusion}
\label{sec:discussion}

We measure the growth index $\gamma$ in a universe with massive neutrinos, through which we make a consistency test for GR. We employ the RSD measurements (eleven data points) to do the analysis. In order to constrain other cosmological parameters, we combine with the Planck 2015 CMB temperature and polarization data, the BAO data, the SN JLA data, the WL galaxy shear data, and the Planck 2015 CMB lensing data. We consider the both cases of sterile neutrino and active neutrino.

In the standard cosmology (the $\Lambda$CDM+$\gamma$ model), we have $\gamma=0.656^{+0.042}_{-0.046}$, with the tension with the standard value of GR prediction $\gamma=0.55$ at the 2.30$\sigma$ level. When massive sterile neutrinos are considered, i.e., in the $\Lambda$CDM+$\gamma$+$m_{\nu,\rm sterile}^{\rm eff}+N_{\rm eff}$ model, we obtain $\gamma=0.624^{+0.055}_{-0.050}$, with the tension relieved to be at the 1.48$\sigma$ level. This result shows that the consideration of massive sterile neutrinos although can lead to a smaller value of $\gamma$, is not capable of making the true measurement of $\gamma$ be well consistent with the GR prediction. In this case, we have $N_{\rm eff}<3.33$ and $m_{\nu,\rm sterile}^{\rm eff}<0.743$ eV.
%, indicating that we find no hint of the existence of light sterile neutrinos in this study.
We also discuss the issue of other tensions of $H_0$ and $S_8$, and make comparison with the latest direct measurement of the Hubble constant and the weak lensing measurements of KiDS-450 and DES Y1.

We also consider the case with massive active neutrinos, and we obtain $\gamma=0.663\pm0.045$ for the $\Lambda$CDM+$\gamma$+$\sum m_{\nu}$ (NH) model, $\gamma=0.661^{+0.044}_{-0.050}$ for the $\Lambda$CDM+$\gamma$+$\sum m_{\nu}$ (DH) model, and $\gamma=0.668^{+0.045}_{-0.051}$ for the $\Lambda$CDM+$\gamma$+$\sum m_{\nu}$ (IH) model. We find that the consideration of massive active neutrinos (no matter what mass hierarchy is considered) almost does not influence the measurement of the growth index $\gamma$. For the three cases, the tensions with the standard value of GR prediction $\gamma=0.55$ are still at the $2.51\sigma$, $2.22\sigma$, and $2.31\sigma$ level, respectively. For the neutrino mass, we have $\sum m_{\nu}<0.390$ eV for the $\Lambda$CDM+$\gamma$+$\sum m_{\nu}$ (NH) model, $\sum m_{\nu}<0.374$ eV for the $\Lambda$CDM+$\gamma$+$\sum m_{\nu}$ (DH) model, and $\sum m_{\nu}<0.406$ eV for the $\Lambda$CDM+$\gamma$+$\sum m_{\nu}$ (IH) model. We find that, in a universe with the density perturbation parametrized by the growth index $\gamma$, the current cosmological observations still cannot diagnose the mass hierarchy of neutrinos.

\acknowledgments
This work was supported by the National Natural Science Foundation of China (Grants No.~11522540 and No.~11690021), the Top-Notch Young Talents Program of China, and the Provincial Department of Education of Liaoning (Grant No.~L2012087).
{}

\end{document}